\documentclass[aip,graphicx]{revtex4-2}

\usepackage{graphicx}
\usepackage{amsmath}
\usepackage{hyperref}

\draft 

\begin{document}


\title{Scanning apparatus to detect the spectral directivity of optically-emissive materials} 



\author{Hengzhou Liu}
\affiliation{Florida Polytechnic University, Lakeland, FL, 33805}

\author{Anthony Fiorito}
\affiliation{Florida Polytechnic University, Lakeland, FL, 33805}

\author{David Sheffield}
\affiliation{Florida Polytechnic University, Lakeland, FL, 33805}

\author{Matthew Knitter}
\affiliation{Florida Polytechnic University, Lakeland, FL, 33805}

\author{Louis Ferreira}
\affiliation{Florida Polytechnic University, Lakeland, FL, 33805}

\author{Nathan Dawson}
\affiliation{Florida Polytechnic University, Lakeland, FL, 33805}


\date{\today}

\begin{abstract}
An apparatus that records the optical spectrum of emissive materials as a function of the polar coordinate angles is reported. The ability for the device to characterize the directive gain of a light source over the optical spectrum is demonstrated. The angular emission profile of an electrically driven LED with a hemispherical diffuser cap was measured. In addition, the device was used to characterize optically pumped materials exhibiting both fluorescence and amplified spontaneous emission (ASE), demonstrating its versatility for diverse emissive systems.
\end{abstract}


\maketitle 

\section{Introduction}

Apparatus designed to detect the directive gain of an emissive material or device are most commonly applied to antennas. Directive gain measurements of transmitting antennae actuate a rotation of the antennae about the azimuth and polar angles while keeping the receiving antenna static.\cite{balan05.01} It is also standard to orient a transponder antennae that can both transmit and receive signals while using a static mirror to bounce the signal from a specific location to determine the directive gain. Anechoic chambers to dampen outside influence are commonplace in antenna frequency ranges.\cite{krchn23.01} For electrically powered optical components, the same antenna directive gain measurement system can be used; however, in optically pumped systems such as materials doped with fluorophores (fluorescent molecules), the pump orientation with respect to the sample should remain static. Therefore, we have designed a system that maintains the sample at static orientation in the lab reference frame while sweeping the polar and azimuthal angles with the detector. Furthermore, materials and devices that emit at optical frequencies often have broad spectral output with combined processes of fluorescence, amplified spontaneous emission (ASE), and laser emission that can be highly directional. Therefore, a directive gain measurement apparatus that scans over the visible spectrum with a miniature spectrometer was developed. The sample and pump beam remains in a fixed position in the lab frame, and the detector sweeps the azimuth and polar angles measuring angular dependent spectra.

This study investigates the directive gain of light emitted from a material for a given sample orientation. Such materials and techniques probe the ``active'' emissive properties of the material. Passive linear optical properties of a material are characterized by detecting changes to an incident probe beam using methods such as reflection, transmission, and absorption spectroscopy.\cite{dyer65.01} Active techniques involve probing the material upon exciting it with i) an external light source,\cite{lakow83.01} ii) electrically-driven charge carriers,\cite{schub06.01} or iii) chemical reactions.\cite{roda11.01} Measuring the directivity over a range of the light spectrum exiting a sample after relaxing from an excited state is the primary focus of this paper.

When an a low population density of excited states of molecular ensembles release electromagnetic energy, the phenomena is referred to as fluorescence.\cite{alban07.01} Dispersing a low concentration of fluorophores in a transparent matrix is the simplest way to ensure a low population density of noninteracting molecules. Fluorescent molecules with long excited-state lifetimes doped in a homogeneous liquid will act as an unpolarized source with the same volumetric shape as the absorbed light. Changing the surface topology of a sample can affect the directivity (\textit{e}.\textit{g}., surface relief grating).\cite{janon21.01,kozan22.01} Furthermore, refractive index-mismatched internal structures can create additional reflecting surfaces which dictate the directivity of light.\cite{manfr17.01}

When a high concentration of fluorophores is pumped by a strong light source, a population inversion density can be achieved. Many fluorescent molecules behave like a 4-level model.\cite{beaum93.01,durko22.01} If the path length of an emitted photon through the gain region of excited molecules is long enough, then amplified spontaneous emission (ASE) can occur,\cite{allen70.01} which is light emitted with a high degree of spatial coherence but without temporal coherence. ASE is easily identified in emission spectra by the nonlinear power dependence of the emission intensity and spectral width. Traditionally, a long strip of material is excited to produce ASE.\cite{peters71.01,rau12.01,stubb18.01,chris19.01} Aperiodic scattering surfaces inside the gain region can also increase the ``effective'' path length of emitted photons which can create directed ASE.\cite{peter11.01,umar19.01,hoink19.01,dawso22.01} The narrow spectral profile for this type of enhanced ASE generation is similar to the spectrum produced by multimode laser emission from an ``intensity feedback'' or ``non-resonant feedback'' random laser.\cite{turec08.01,ignes13.01,ander16.02,moura23.01} Natural and fabricated materials can possess a highly periodic internal structure.\cite{sun13.01,elias24.01,sheff24.01,komik07.01,song09.01,dawso23.01} When the internal structure is highly periodic, laser light can be emitted (\textit{e}.\textit{g}., multilayer distributed feedback laser).\cite{kogel71.01,andre12.01,andre13.01,golde15.01} At high thresholds, even aperiodic structures can produce laser emission, which has been referred to as a ``resonant feedback'' random laser.\cite{conso15.01,biasc19.01,hara24.01} The directive gain from the complex structures and properties given in the latter examples can be probed by measuring the spectrum of the emitted light over the polar and azimuth angles. This paper presents a spectral scanner, which collects light via a miniature spectrometer while scanning through the polar and azimuth angles to characterize the emissive properties of simple and complex-structured materials and devices.

\section{Preliminary theory}
\label{sub:specdir}

To understand scan results from the apparatus, it may be beneficial to first discuss the directive gain of a radiative source. The time-averaged power density (the intensity) vector of a monochromatic wave can be determined from the Poynting theorem,
\begin{equation}
\vec{I} \left(r,\phi,\theta\right) = \frac{1}{2}\mathrm{Re}\left[\vec{E}_\mathrm{s} \times \vec{H}_{\mathrm{s}}^\ast\right]\, ,
\label{eq:pavg}
\end{equation}
with the asterisk denoting the complex conjugate. We have expressed the electric field $\vec{E}_\mathrm{s}$ and magnetic field $\vec{H}_{\mathrm{s}}$ in the usual phasor form for a monochromatic wave of angular frequency $\omega$, where
\begin{align}
\vec{E}\left(r,\phi,\theta,t\right) &= \vec{E}\left(r,\phi,\theta\right) \mathrm{Re}\left[e^{i\left(\omega t + \varphi\right)}\right] \nonumber \\
&= \mathrm{Re}\left[\vec{E}_\mathrm{s} e^{i \omega t}\right] \, .
\label{eq:elecvec}
\end{align}
Here, the field $\vec{E}_\mathrm{s}$ in phasor form must follow as
\begin{equation}
\vec{E}_\mathrm{s} = \vec{E}\left(r,\phi,\theta\right)\mathrm{Re}\left[ e^{i\varphi}\right]\, ,
\label{eq:esdef}
\end{equation}
where $\varphi$ is a constant that relates the initial phase of the wave when $t = 0$.

In the far-field, the radiated power is directed radially outward from the source, $\vec{I} \left(r,\phi,\theta\right) = I \left(r,\phi,\theta\right)\hat{r}$. Therefore, alignment of the sample to be the same distance away from the detector at all scanned angles is crucial to the accuracy of the scan. The average power radiated is found by integrating the intensity by the area of a sphere centered at the source in the far field. The intensity for some small solid angle patch is maximized for some set of spherical coordinate angles, $I_\mathrm{max}$. The normalized radiation intensity as a function of the spherical polar coordinates is expressed as
\begin{equation}
I_\mathrm{norm}\left(\phi,\theta\right) = \frac{I\left(\phi,\theta\right)}{I_\mathrm{max}} \, .
\label{eq:norm}
\end{equation}

It is often useful to find the average value for the normalized radiation intensity over an entire sphere with the source at the center. In this case,
\begin{equation}
I_\mathrm{norm,avg} = \displaystyle \frac{\int I_\mathrm{norm}\left(\phi,\theta\right) \,d\Omega} {\int d\Omega} \, ,
\label{eq:normsph}
\end{equation}
where $d\Omega = d\phi\,d\left(\cos\theta\right)$ is the solid angle. The denominator in Equation \ref{eq:normsph} is $4\pi$. For a monochromatic source such as an antenna, the directive gain is determined by the ratio of the normalized radiation intensity and the average normalized radiation intensity integrated over an entire sphere,
\begin{equation}
D\left(\phi,\theta\right) = \displaystyle \frac{I_\mathrm{norm}\left(\phi,\theta\right)} {I_\mathrm{norm,avg}} \, .
\label{eq:direc}
\end{equation}
The directivity is defined as the maximum directive gain. A simple way to determine the directivity of a source emitting a monochromatic wave is by first determining the pattern solid angle,
\begin{equation}
\Omega_\mathrm{pattern} = \int I_\mathrm{norm}\left(\phi,\theta\right) \,d\Omega \, .
\label{eq:pattern}
\end{equation}
Note that Equation \ref{eq:pattern} allows the average normalized radiation intensity integrated over the entire sphere to be rewritten, where $I_\mathrm{norm,avg} = \Omega_\mathrm{pattern}/4\pi$. From the normalization condition, it also follows that the maximum value of $I_\mathrm{norm}\left(\phi,\theta\right)$ is 1. Therefore, the directivity, which is the maximum value of the directive gain, can be easily expressed in terms of the pattern solid angle,
\begin{equation}
D_\mathrm{max} = \frac{4\pi} {\Omega_\mathrm{pattern}} \, .
\label{eq:direc}
\end{equation}

As written in Equation \ref{eq:direc}, the directive gain is defined for monochromatic sources. Many visible light sources have a broad emission bandwidth. Although the data is typically graphed with the vertical axis labeled as ``intensity,'' spectrometers typically measure the intensity spectral density, $S$, which has dimensions of intensity per unit frequency or per unit wavelength. Integrating the spectral curve yields the total intensity of light, \textit{i}.\textit{e}., $I = \int S\,d\lambda$. For a spectrometer measuring the intensity spectral density over the spherical angles, the stored data is in the form of a hypercube. This requirement is akin to storing hyperspectral image data that stores the Cartesian ($x,y$) image positions as well as the multiple wavelength channels.

We introduce the wavelength-dependent, spectral directive gain density, ${\cal D}_\lambda \left(\phi,\theta,\lambda\right)$. Integrating the wavelength-dependent, spectral directive gain density over all wavelengths yields the directive gain $D\left(\phi,\theta\right)$ for the total power emitted over the entire spectrum,
\begin{equation}
D\left(\phi,\theta\right) = \int {\cal D}_\lambda \left(\phi,\theta,\lambda\right) \,d\lambda
\label{eq:direcS}
\end{equation}
The normalization condition requires that the wavelength-dependent, spectral directive gain density be defined over the wavelength range of the normalized radiation intensity spectral density,
\begin{equation}
{\cal D}_\lambda \left(\phi,\theta,\lambda\right) = \displaystyle \frac{S_\mathrm{norm}\left(\phi,\theta,\lambda\right)} {\int S_\mathrm{norm,avg}\left(\lambda\right) \,d\lambda} \, .
\label{eq:direcnormS}
\end{equation}
Note that $S_\mathrm{norm}\left(\phi,\theta,\lambda\right)$ is normalized over the total spectrum, where the largest value of $S_\mathrm{norm}\left(\phi,\theta,\lambda\right)$ is not equal to 1, which was the case for $I_\mathrm{norm}\left(\phi,\theta\right)$. The normalization condition results in $I_\mathrm{norm}\left(\phi,\theta\right)$ after integration of $S_\mathrm{norm}\left(\phi,\theta,\lambda\right)$ over the wavelength spectrum.
The average normalized radiation intensity spectral density integrated over the sphere is given by
\begin{equation}
S_\mathrm{sph,norm}\left(\lambda\right) = \displaystyle \frac{\int S_\mathrm{norm}\left(\phi,\theta,\lambda\right) \,d\Omega} {\int d\Omega} \, .
\label{eq:normsphS}
\end{equation}

It is often difficult for experimental data to capture the full range of a theoretical concept. For example, no single detector can be used for all possible frequencies ranging from radio frequency to x-rays. Therefore, measurement data must account for these experimental limitations. For externally pumped samples, the detector cannot be allowed to block the pump beam. Therefore, low co-latitude angles are not always available to scan. Likewise, the apparatus presented in this article has a sample cooling stage, which is useful for biological materials that can degrade at a higher rate at room temperature. Therefore, the sample mounting technique will also affect the range over which the emission can be scanned. The average normalized radiation intensity spectral density integrated over the sphere given in Equation \ref{eq:normsphS} cannot be divided by the solid angle of a full sphere if the scanning apparatus could not scan for light. In some cases, the intensity can be assumed to be quite low in some directions such as when a sample is mounted on a mirrored surface. Other cases may indeed allow for much light to be emitted, but unable to detect with the apparatus. Therefore, we must consider a worthy measure over the scanned surface.

Let us consider the integral in the denominator of Equation \ref{eq:normsphS}. If the scanned area is not over the full sphere, then we should consider normalizing the function over the scanned area only. Then substituting the normalized radiation intensity spectral density integrated over the scanned area only, we arrive at a metric useful for incomplete scans, which is the scanned-area, spectral directive gain density,
\begin{equation}
{\cal D}_{\lambda,\mathrm{SA}} \left(\phi,\theta,\lambda\right) = \displaystyle \frac{S_\mathrm{norm}\left(\phi,\theta,\lambda\right)} {\int S_\mathrm{norm,avg,\mathrm{SA}}\left(\lambda\right) \,d\lambda} \, ,
\label{eq:direcnormSAS}
\end{equation}
where the denominator is averaged over integral limits associated with the angular patch of the sphere that was scanned by the apparatus. Integrating ${\cal D}_{\lambda,\mathrm{SA}} \left(\phi,\theta,\lambda\right)$ over all wavelengths yields the scanned-area, directive gain, $D_{\mathrm{SA}} \left(\phi,\theta\right)$.

Note that the total power emitted over the entire spectrum can be useful to find hot spots in light sources that are relatively unremarkable in their spectrum as a function of angle. Examples of such light sources are light-emitting diodes (LED), incandescent light bulbs, and thin films doped with a low-concentration of fluorophores. The apparatus presented in this article collects raw spectral data, which can be used to determine the spectral directive gain. The alignment is characterized by generating visual map of the directive gain while the wavelength-dependent, spectral directive gain is visualized at specific angles to study the dominance of different emission processes directed at different angles.

\section{Experimental methods}
\label{sec:methods}

\subsection{Apparatus}
\label{sec:app}

\begin{figure}[t]
\centering\includegraphics[width=12cm]{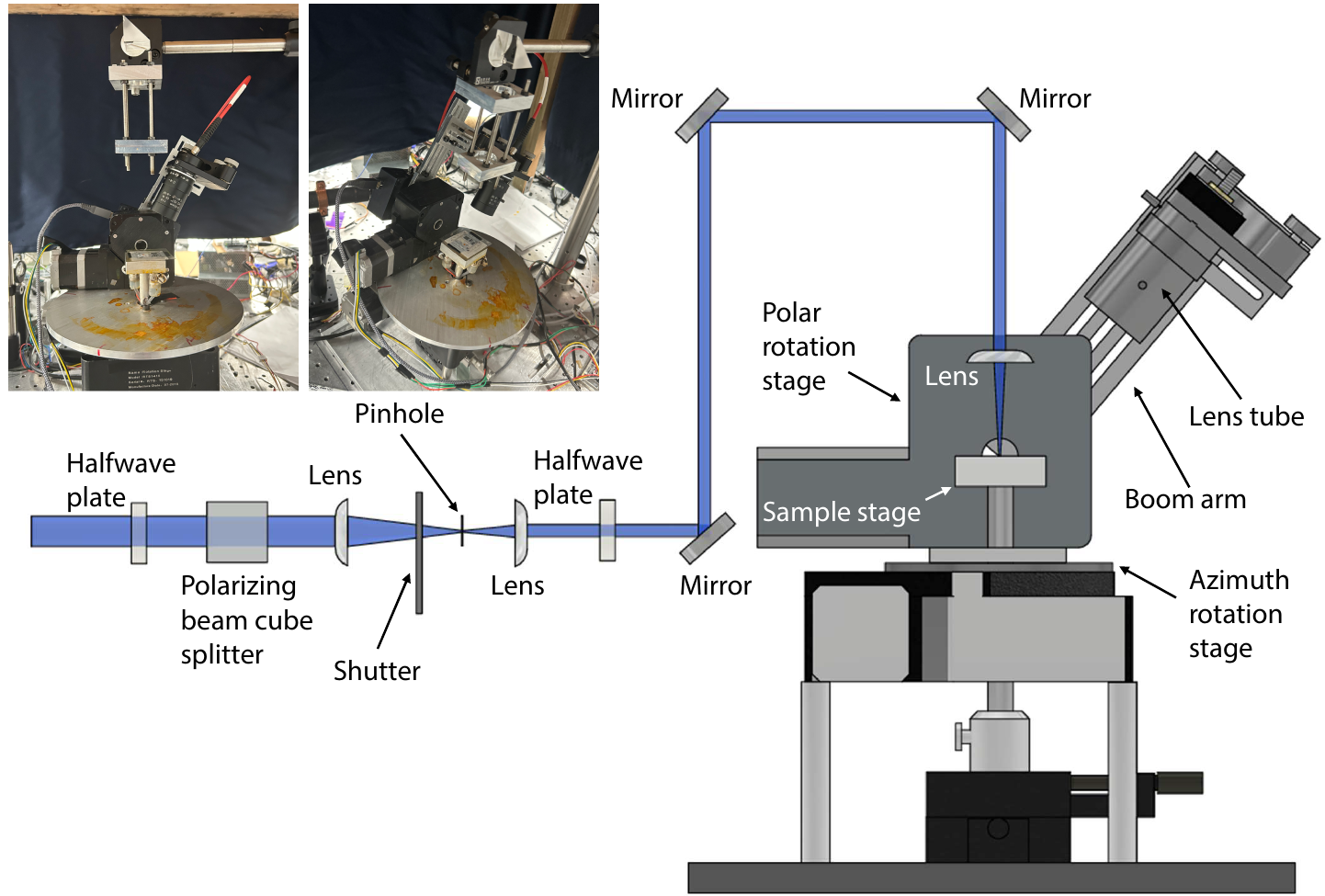}
\caption{Labeled illustration of apparatus used to spherically scan the spectrum of an emissive material. The two inset images in the upper left corners show the built device from two different perspectives.}
\label{fig:fig1}
\end{figure}

Automated recordings of emission spectra scanned over the azimuth and polar angles were performed by an in-house built apparatus. An aluminum sample stage was at the center of the apparatus and temperature controlled by running liquid through a bored cavity. The sample stage was secured by a post that ran through a bore hole in a rotation stage. The sample stage remained in a fixed position while the rotation stage was allowed to sweep the azimuth angle as shown in Figure \ref{fig:fig1}. A circular plate was then attached to the azimuth rotation stage. An off-center rotation stage was attached to the circular plate so that it could sweep the polar angles. The circular plate was used to mark rotation angles to initially configure the azimuth rotation stage at the speed of $14\,^\circ$/s. A boom arm was attached to the second rotation stage which swept the polar angle. The polar angle was displaced between two positions at a speed of $7\,^\circ$/s. A lever arm was attached at the top of the boom arm to secure the detector and its alignment components.

A tunable double lens tube was used to zoom and focus in on the sample stage. Adjustments to the lens tube's orientation were made by mounting it on a kinematic mirror mount. The kinematic mirror mount is mounted on a slot and the distance between zoom lens tube and the sample plate can be adjusted, A camera was placed in the image location for alignment purposes. After alignment, the camera was replaced by a $800\,\mu$m-core multimode optical patch fiber. The fiber was connected to a portable spectrometer mounted to the boom arm. We initially attempted to secure one end of a long fiber to the apparatus and the other end to a spectrometer mounted on the optical table, but fiber deformations from apparatus rotations caused the relative intensity of the signal to change. By mounting the spectrometer to the boom arm with a rigid fiber patch cord, loss through the optical fiber was held constant. A slot was located between the lens tube and the detector position which allowed for 1" filters (chromatic, polarization, etc.).

The motors on the apparatus were controlled by a CNC shield attached to an Arduino Uno loaded with a GRBL library. The Arduino was connected to a computer via USB port. The spectrometer was also connected to the computer. A Python script recorded spectra with the Seabreeze package, and it also controlled the apparatus motors. A third motor was connected to the CNC shield which rotated a half-wave plate as shown in Figure \ref{fig:fig1}. The pump beam's polarization was rotated and sent through a polarizing beam cube splitter. Pulse energy was adjusted using this half-wave plate and polarizer combination. A relay was also connected by USB port and incorporated into the same Python code that controlled the entire apparatus. The relay controlled a shutter, where the shutter was open for data collection of materials excited by the external pump laser and closed when the apparatus rotated during the scanning procedure. The shutter reduced fluorophore photodegradation during a scan.

\subsection{Alignment}
\label{sec:align}

The sample stage can be placed at the approximate center of the azimuth rotating stage bore hole using $xy$-positioning stage at the base of the apparatus. The sample stage height can be adjusted so the sample plate is placed approximately $2-3$ $mm$ lower than center of the polar rotating stage bore hole. Later on, the center of the azimuth rotation stage on the sample plate can be found and the sample will be placed on the spot.

Before connecting the optical fiber, a camera was first positioned behind the adjustable lens tube. The camera was used to finely align the system via an iterative technique. First the lens system has to be aligned with the arm. Placing the camera at a polar angle of $90^\circ$ and rotating the azimuth plate, the center spot of the azimuth rotation stage can be found. We then marked the spot on the computer screen. We noticed that the sample plate may not be perfectly at the center of the azimuth rotation stage. Using $xy$-positioning stage at the base of the apparatus to move the sample plate and then marked the center of the rotation stage on the sample plate. This spot is the sample position which will be aligned with the center of the camera. We can adjust the camera position so the sample position is in the center of the camera. After that, the apparatus then swung around $360^\circ$. If the sample position rotated in a small circle or didn't move, the kinematic mount was adjusted until the sample remained in the same position through the entire scan.

After the sample position was aligned in $xy$-plane. A LED or a droplet (sample) can be placed in the center to continue next alignment step. The detector was placed at a polar angle of $0^\circ$. Then we adjusted the sample plate height to place the top of the sample in the center of the camera. Then the apparatus swung around $360^\circ$ and if the sample didn't move or move in a line, then we can use the kinematic mount. A final check was performed by scanning the entire set of azimuth and polar angles to confirm the sample position remained constant from the camera perspective. The camera was then replaced by the fiber feeding into the spectrometer attached to the boom arm.

\subsection{Apparatus testing and experimentation}
\label{sec:exp}

The apparatus measures the directive gain of an emissive material/device over the spectral range recorded by the fiber spectrometer. Two types of samples were recorded, 1) an electrically-driven light emitting diode (LED) with a diffuser cap and 2) a fluorescein-doped water droplet. The electrical current for the LED and the pump beam pulse energy for the fluorescein-doped water were held constant for the duration of their scans. The droplets were carefully measured before each scan and after each scan using a plastic scale that was placed underneath the droplet. All scans were stepped $10^\circ$ in the azimuth from $0^\circ$ to $350^\circ$. They scans were stepped $10^\circ$ along the co-latitude from $20^\circ$ to $70^\circ$.

The two different sizes of droplet samples were measured under the same geometric condition. For the first one, a large dye-doped water droplet was placed on a parafilm surface with a pump beam illuminating the sample from above. The hydrophobic surface caused the droplet to form a dome shape. A second droplet's diameter was approximately half the diameter of the first droplet and pumped with a similar size pump beam. A $470\,$nm-wavelength, $7\,$ns pulsed laser beam operating at $10\,$Hz was used to pump the liquid droplets. The wavelength was chosen as $470\,$nm because it is highly absorbed by fluorescein while also absorbed by GG475 Schott glass. The GG475 filter placed in the detector optics chain passed all detectable fluorescence while blocking stray pump light that can saturate spectrometer pixels.

\section{Results}
\label{sec:results}

\subsection{Diffuser-capped LED}
\label{sub:led}

Our system was aligned through the process described in Section \ref{sec:align}. In an attempt to confirm the alignment, an LED with a diffuser cap was placed at the center of the sample stage. An ideal spherical diffuser cap would mimic a spherically symmetric light source, which would confirm the alignment; however, the spherical diffuser capping the LED did not have spherical light-emissive symmetry. Although the normalized spectral profile was independent of angle, the output intensity had angular dependencies along both the polar and azimuth directions.

\begin{figure}[h]
\centering\includegraphics[width=12cm]{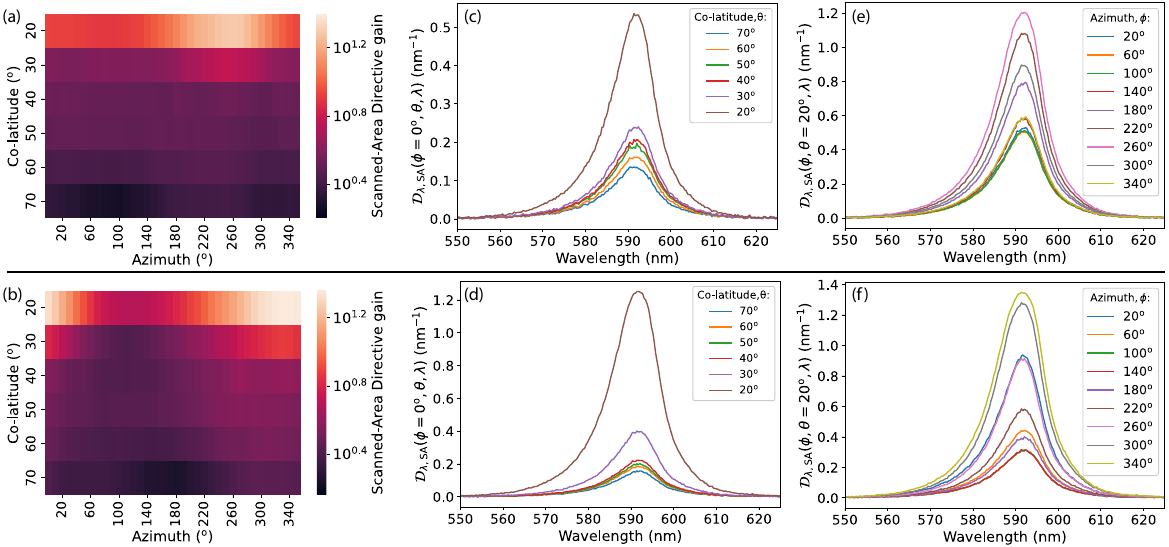}
\caption{A heat map showing the scanned-area directive gain as a function of azimuth and co-latitude (spherical polar coordinate) angles for (a) the initial LED orientation and (b) the LED rotated in the horizontal plane by $\sim 90^\circ$. The LED's emission spectrum as a function of detector co-latitude positions at a fixed azimuth for (c) the initial LED orientation and (d) the rotated LED. The emission spectrum as a function of detector's azimuth angle at a fixed co-latitude for (e) the initial orientation and (f) the rotated orientation.}
\label{fig:fig2}
\end{figure}

At high polar angles, a hot spot was formed along the azimuth as depicted in the heat map representing the scanned-area, spectral directive gain density given in Figure \ref{fig:fig2}(a). The nonuniform $D_{\mathrm{SA}} \left(\phi,\theta\right)$ could have been caused by an alignment issue, so the sample was rotated by $\sim 90^\circ$ and a new scan was performed. As shown in Figure \ref{fig:fig2}(b), the recorded hot spot rotated in the azimuthal direction by the same angle that the sample was rotated. This azimuth rotation of the sample confirmed the hot spot was indeed sourced from the diffuser-capped LED as opposed to an issue with the system alignment.

As expected, there was no distinguishable difference between the peak wavelength and spectral width of the LED's emitted light. The amplitude of the spectra shown in Figure \ref{fig:fig2}(c-f) vary in amplitude only.  There is a significant trend for the ${\cal D}_{\lambda,\mathrm{SA}} \left(\phi,\theta,\lambda\right)$ dependence on the polar angle as shown in Figure \ref{fig:fig2}(c-d). The detected signal is largest near the top of the LED and significantly weakens as the detector collects light near the horizon. A hot spot is clearly identifiable as a function of the azimuthal angle, the peak amplitude of the spectra, represented by ${\cal D}_{\lambda,\mathrm{SA}} \left(\phi,\theta,\lambda\right)$ and shown in Figure \ref{fig:fig2}(e-f), shifted by the same angle as the rotated sample. The LED has no significant process for emitting amplified light, and the diffuser cap is fabricated from a passive achromatic material. Therefore, as expected for a linear optical source, the spectral features remain constant while only the amplitude of the emitted light depends on the spherical angles. This result implies that highly directed energy is emitted from the source or during the passive diffusion through the diffuser cap, as expected.

\subsection{Large water droplet: center pumped}
\label{sub:large}

The diffuser-capped, electrically-pumped, LED light source in the previous section had hot spots caused by aspherical light emission. A $1\,$wt.\% fluorescein/water solution was also prepared in an attempt to create a better approximation to a spherical-emitting light source. The droplet was placed on parafilm, where the solution/air interface above the parafilm surface formed an dome that can be approximated as a hemisphere. The beam spot was $\sim 450\,$$\mu m$ in diameter, which illuminated the majority of the liquid droplet's surface. The liquid cooled stage was measured to be $\sim 4\,^\circ$C while the ambient air was held at room temperature and moderate humidity. The cooled surface resulted in a near steady state droplet size due to the competition between evaporation and condensation.

\begin{figure}[t]
\centering\includegraphics[width=12cm]{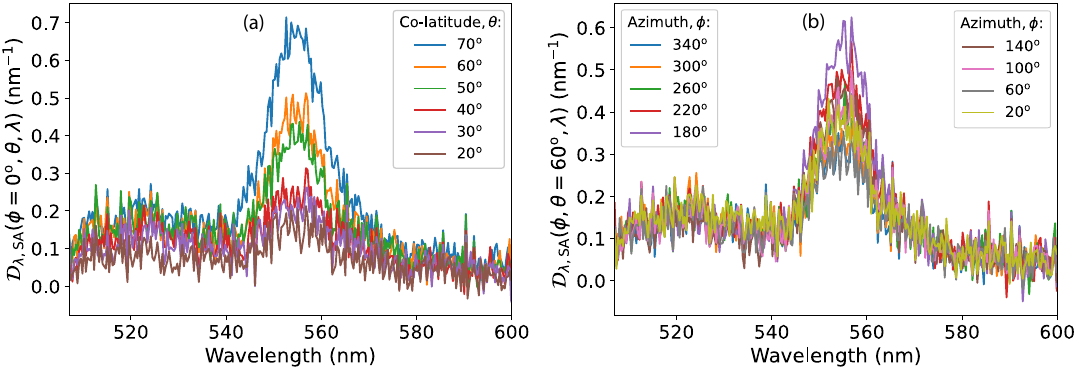}
\caption{The emission spectra as functions of (a) co-latitude at a fixed azimuth and (b) azimuth angle at fixed co-latitude for a droplet of fluorescein pumped from a vertical beam from an OPO tuned to $470\,$nm.}
\label{fig:fig3}
\end{figure}

There is an increased emission intensity at higher co-latitude (lower latitude) for the droplet. This trend in emission as a function of polar angle is opposite of the diffuser-capped LED, where the scanned-area, spectral directive gain density shown in Figure \ref{fig:fig3}(a) can be visually compared to those given in Figure \ref{fig:fig2}(c).

The noise associated with the spectrum is also greater in the droplet at low emission intensity levels as compared to the diffuser-capped LED. The azimuthal dependence shown in Figure \ref{fig:fig2}(b) also has some shot-to-shot variations throughout the scan. Some small variations in the beam profile could result in an azimuth-dependent emission; however,noise from the pump beam also contributes to the variations in the azimuthal dependence of the emission intensity. The shot-to-shot noise of an Nd:YAG's fundamental frequency is reasonable low. The shot-to-shot uncertainty of the pump is increased when a portion of the $1064\,$nm pump beam's photons are frequency doubled by a beta Barium Borate (BBO) crystal followed by sum-frequency generation of the $1064\,$nm and $532\,$nm to yield a $355\,$nm beam. This nonlinear, macroscopic, cascading process results in larger variations between pulse energies for the $355\,$nm beam entering an optical parametric oscillator (OPO). Prior to the pump light entering the OPO cavity, the residual $1064\,$nm and $532\,$nm light is attenuated by a set of dichroic mirrors. The remaining $355\,$nm light is injected into a cavity \textit{via} another dichroic mirror with a long BBO crystal between a high reflectance mirror and a dichroic output coupler. The amplified output of the OPO for wavelength selection at $470\,$nm also amplifies the shot-to-shot uncertainty of the pulse energy. Thus, the uncertainty between each shot appears to be relatively large.

Averaging out the uncertainty between recorded shots taken while the azimuth is scanned, a slow change of intensity is observed through the scan as shown in Figure \ref{fig:fig3}(b). Therefore, the calculated ${\cal D}_{\lambda,\mathrm{SA}} \left(\phi,\theta,\lambda\right)$ shows some preference for an azimuthal emission direction. If the preferred direction were biaxial, the cause would undoubtedly be from the pump beam's polarization. The excited state lifetime of fluorescein is relatively long, making a polarization dependent emission unlikely in water. Indeed, we found that the small, preferred, azimuth emission was unidirectional, which was caused by a small uncertainty in both the pump beam profile and center-alignment. The ASE clearly dominates the fluorescence signal. The small fluorescence peak is observed at a $\sim 520\,$nm wavelength while the peak of the red-shifted ASE profile is much greater at $\sim 555\,$nm.

\subsection{Small water droplet: off-center pumped}
\label{sub:small}

\begin{figure}[t]
\centering\includegraphics[width=12cm]{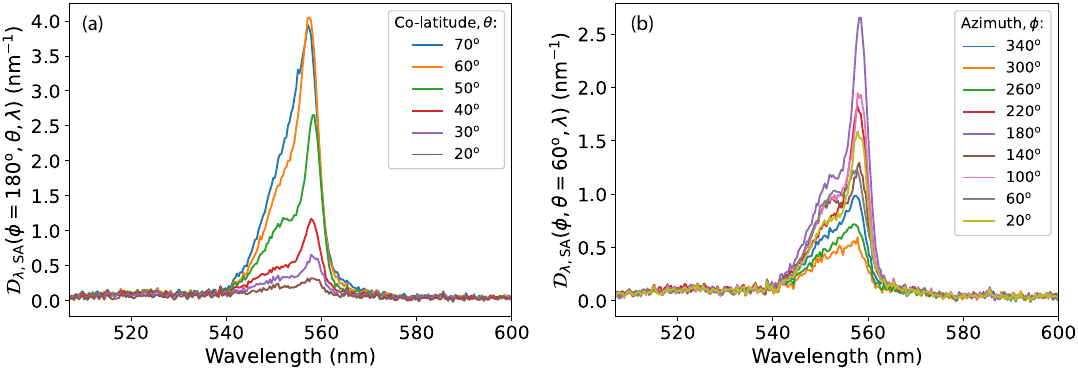}
\caption{Emission spectra of an asymmetrically pumped liquid droplet as functions of (b) co-latitude at a fixed azimuth and (c) azimuth angle at fixed co-latitude.}
\label{fig:fig4}
\end{figure}

The spectral directive gain density of the emission spectra for a smaller droplet of $1\,$wt.\% fluorescein/water resting on parafilm was also scanned. The droplet in Section \ref{sub:large} had twice the diameter of the smaller droplet which equates to four times the cross-sectional area. The pump beam spot was the same size for both droplets and its diameter was measured to be $\sim 430\,\mu$m. The pump pulse energy and compact gain region resulted in an above threshold mixture of ASE and fluorescence. Similar to the large droplet described in Section \ref{sub:large}, the spectral directive gain density increased with the co-latitude as shown in Figure \ref{fig:fig4}(a).

The focused beam was slightly off center in this case which caused an asymmetric emission profile as measured in the azimuthal direction. Figure \ref{fig:fig4}(b) shows emission spectra as a function of the azimuth angle for a fixed $60^\circ$ co-latitude. The small deviation from the vertical axis centered on the hemispherical droplet had significant effects on the azimuthal dependence of the emission intensity and type of radiative process. The fluorescence and ASE emission intensities were comparable when viewed from an azimuth angle of $\sim 300^\circ$. As the detector scanned around the sample, the ASE began to dominate the emitted light signal. As with the large droplet, there was a significant amount of uncertainty caused by the shot-to-shot deviations of the OPO pulse energy. Even after considering this uncertainty, there is a clear trend associated with the spectral directive gain density and the azimuthal angle.

\section{Discussion}

The diffuser-capped LED was initially intended to confirm alignment based on an assumed spherical symmetry for light emission from cosine correction across the hemispherical diffuser cap. In hindsight, the polar angle dependence can be easily explained when assuming a highly directional LED light source directed towards the top of the diffuser cap from the inside. The hot spot identified in the heat map for the scanned-area, spectral directive gain density was shown to be caused by the source as opposed to an error in alignment of the apparatus. Additional rotations beyond those shown in Figures \ref{fig:fig2}(a) and \ref{fig:fig2}(b) further confirmed the LED source had an azimuthal dependence on the emission intensity. Thus, the apparatus can be used to characterize electrically-driven light sources in addition to its initially intended purpose of characterizing gain-doped materials with complex structure functions.

The high dye concentration in the water droplets was chosen because of the sensitivity of ASE emission. The ASE process occurs when 1) a population inversion density is present and 2) the path length of an emitted beam is long enough to amplify the emitted light. The amplification of spontaneous emission when the OPO-pumped droplet was nearly cylindrically symmetric resulted in a near azimuthal symmetry for the emission; however, the uncertainty caused by shot-to-shot intensity variations of the OPO pump introduced some noise-like variance between each recorded spectrum which can be observed in Figure \ref{fig:fig3}(a). The directive gain was found to be independent of pump polarization.

The characterization for the droplet that was pumped off-center provided interesting insights. One might expect the ASE and fluorescence intensities to be weighted evenly with respect to each other at all emission angles. Instead we observed that the directive gains associated with the two emission processes were different from each other as seen in the scanned-area, spectral directive gain density shown in Figure \ref{fig:fig4}. The results from this last study also illustrate the importance of pump beam characteristics in addition to apparatus alignment when scanning optically pumped light sources.

Although the apparatus was initially fabricated for use with a miniature spectrometer, other detectors could take its place. For example, a fast photodiode could be mounted to collect transient data, where emitted pulses differ between fluorescence, ASE, and lasing. This was not attempted with the current setup, where there was already a large uncertainty in the OPO's shot-to-shot pulse energy appearing with spectrometer data. It may be of interest to test other sources such as a rhodamine derivative solution pumped by a frequency-doubled Nd:YAG, which should result in lower pulse energy deviations.

Additional perturbations on the original design are also possible. For example, the reflected and scattered light intensity as a function of wavelength and polar coordinate angles can be measured instead of the emission of a gain-doped sample. In this alternative design, a slewing bearing can be motorized in addition to the plate attached to the borehole motor which allows for a source (such as a collimated flat-spectrum LED) and a detector to rotate in the azimuthal direction independently. The source and detector can then have two separate motors to adjust the polar angle for each one. From such an alternative design, the reflected and scattered light could be measured as a function of source position, detector position, and wavelength, \textit{i}.\textit{e}., the detected light intensity would follow as $I\left(\phi_\mathrm{source}, \theta_\mathrm{source}, \phi_\mathrm{detector}, \theta_\mathrm{detector}, \lambda \right)$.

\section{Conclusion}
\label{sec:conc}

This article describes an apparatus that scans the spectrum of emitted light as a function of the spherical coordinate angles. Alignment of the apparatus was achieved through the use of a camera which imaged the light source as it swept through a series of angles. The apparatus was found to be well-suited for characterization of LED light sources as well as emissive materials that require optical pumping. Beyond the example materials characterized in Section \ref{sec:results}, the scanned-area, spectral directive gain density of many types of emissive materials can be characterized. Additional examples include random lasers, light sources surrounded by dispersive elements, and complex biological materials stained with fluorescent molecules.

\vspace{0.5cm}

\noindent \textbf{Funding:} This material is based upon work supported by the National Science Foundation, Directorate for Mathematical and Physical Sciences (Grant No. 2337595).

\vspace{0.5cm}

\noindent \textbf{Acknowledgment:} The authors thank Mike Kalman for his role as machinist during the build of the apparatus described in this paper.

\vspace{0.5cm}

\noindent \textbf{Disclosures:} The presented device was granted a US Patent 12203850 on Jan 21 2025.

\vspace{0.5cm}

\noindent \textbf{Data availability:} Data underlying the results presented in this paper are not publicly available at this time but may be obtained from the authors upon reasonable request.


\end{document}